\documentclass[aps,twocolumn,twoside,groupedaddress]{revtex4}
\usepackage{graphicx}
\usepackage{epsfig}
\usepackage{epsf}
\usepackage{amssymb}
\usepackage{amsmath}
\usepackage[toc]{appendix}
\usepackage[caption=false]{subfig}

\begin{document}

\title{Quantum state preparation by a shaped photon pulse in one-dimensional continuum}

\author{Zeyang Liao\footnote{zeyangliao@physics.tamu.edu} and M. Suhail Zubairy }

\affiliation{Institute for Quantum Science and Engineering (IQSE) and Department of Physics and Astronomy, Texas A$\&$M University, College Station, TX 77843-4242, USA 
}

\begin{abstract}
We propose a method to deterministically prepare a desired quantum state in a one-dimensional (1D) continuum by a shaped photon pulse. This method is based on time-reverse of the quantum emission process. We show that the desired quantum state such as Dicke or timed-Dicke state can be successfully prepared with very high fidelity even if the dissipation to the environment is nonnegligible and the pulse shaping is not perfect. We also show that large quantum entanglement between emitters can be created by just a single photon pulse. This method is experimentally feasible in 1D waveguide-QED or circuit-QED system.  
\end{abstract}

\maketitle

\section{Introduction}

Preparation of a quantum state such as highly entangled state has important applications in quantum information, quantum simulation, and quantum metrology \cite{Nielsen2000}. One of the most widely used methods to prepare a quantum state is by applying a seriers of unitary quantum gates to drive the system into the diresed state \cite{Law1996, Di2004, Lukin2000, Leifer2003, Song2005, Hofheinz2009, Ladd2010}. The realization of the quantum gates is significantly restricted by the decoherence time of the system. An alternative way to prepare a quantum state is via bath environment engineering. This method may dissipatively drive a quantum system to a desired state without worrying about the decoherence of the system \cite{Kraus2004, Krauter2011, Ma2013, Lin2013, Aron2014, Gonzalez-Tudela2015, Ma2015, Aron2016}. However, the design of quantum gates and bath environment is usually complicated especially when the many-body interaction is present.

In this paper, we propose a more straightforward method to prepare a quantum state by a shaped photon pulse which is differnet from the methods based on quantum gates.  This method is based on the time reversal symmetry of a closed quantum system \cite{Sakurai2014, Crooks2008, Yuan2016}. More specifically, a desired quantum state can be prepared by inverting a quantum emission process.  It is known that the dynamics of a quantum system is governed by the Schr\"{o}dinger equation, i.e., $i\hbar \partial /\partial_{t} |\psi(t)\rangle = H(t) |\psi(t)\rangle$.  By applying complex conjugate on both sides and taking $t\rightarrow -t$, we have $i\hbar \partial /\partial_{t} |\psi^{*}(-t)\rangle = H^{*}(-t) |\psi^{*}(-t)\rangle$. Hence, complex conjugate of the quantum state with reverse time also satisfies the Schr\"{o}dinger equation govern by $H^{*}(-t)$. If $|\psi(0)\rangle \xrightarrow {H(t)} |\psi(T)\rangle$, then we also have
$|\psi^{*}(T)\rangle \xrightarrow{H^{*}(-t)} |\psi^{*}(0)\rangle$. To prepare a quantum state $|\psi(T)\rangle$ in a multi-emitter system, we can input a photon pulse with spectrum being the complex conjugate of the emission spectrum of the same quantum system prepared in the quantum state $|\psi^{*}(T)\rangle$. Hence, the design of required photon pulse is straightforward and different from previous methods the photon number required to prepare a quantum state in current method is minimized.

In the usual three-dimensional space, the time-reversal of an emission system is almost impratical because the photon is emitted to all directions. In contrast, the photon only emits to two directions (left or right) in the 1D waveguide-QED system \cite{Shen2005, Zheng2013b, Shi2015, Douglas2015, Liao2016, Konyk2016, Roy2017, Gu2017}. It is therefore more feasible to reverse the emission process and prepare a desired quantum state in this system. It has been shown that full inversion of a two-level emitter and high effcient quantum state transfer between two emitters are possible in a 1D continumn by specially designed photon pulse \cite{Rephaeli2010, Wang2011, Pinotsi2008, Korotkov2011, Srinivasan2014}. Here, we show a general procedure to prepare a quantum state in a multi-emitter system coupled to a 1D structure with many-body interactions inlcuded. To illustrate our method, here for simplicity we mainly consider the single-excitation case which has analytical solution and the waveguide is assumed to be a pure 1D waveguide model which is valid when the emitters mainly couple to a single mode of a quasi-1D waveguide like a line defect in a photonic crystal \cite{Englund2007} or superconduting transmission line \cite{Wallraff2004, Hoi2011}. We also consider the noises like decay to the free space, imperfect pulse shaping and emitter position uncertainty and our numerical simulation shows that our method is robust against these noises.

This paper is organized as follows. In Sec. II, we illustrate the schematic setup and basic principle for preparing a single-excitation quantum state. In Sec. III, we use the numerical results to show how our method works. In Sec. IV, we discuss how to prepare a robust quantum state against decoherence. Finally, we summarize the result. 
   
\begin{figure}
\includegraphics[width=0.9\columnwidth]{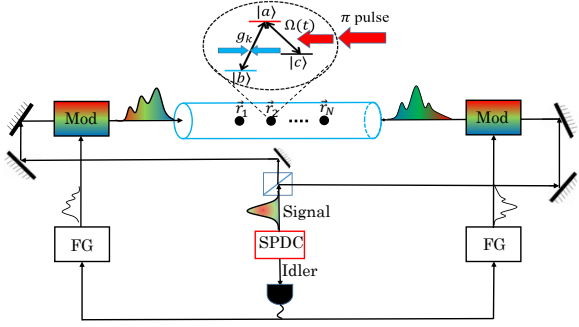}
\caption{(Color online) The schematic setup to generate a single-excitation quantum state in a 1D waveguide-QED system. The single photon source can be generated by the spontaneous down conversion (SPDC) process.  The idler photon can trigger the function generator (FG) to control the photon modulator. The photon modulator (Mod) can shape the signal photon to the desired shape. The shaped single photon pulse is injected from both ends of the waveguide to excite the $|a\rangle\leftrightarrow |b\rangle$ transition and a classical $\pi$ pulse is applied to transfer the population from $|a\rangle$ state to $|c\rangle$ state.  } 
\end{figure}

\section{Schematic setup  for quantum state preparation in 1D space}

The schematic setup is shown in Fig. 1 where $N$ emitters with positions $\mathbf{r}_1,\mathbf{r}_{2},\cdots, \mathbf{r}_{N}$ are coupled to a 1D waveguide. The atoms are assumed to be identical and they have $\Lambda$-type energy structure. The waveguide photon mode can couple to the $|a\rangle\leftrightarrow |b\rangle$ transition, while $|a\rangle\leftrightarrow |c\rangle$ transition is driven by a classical light pulse. It is assumed that $|c\rangle$ state is substable with very slow decay rate. To prepare a desired quantum state, we first calculate the required photon pulse based on the time-reversal process. Then we can generate the required photon pulse by certain pulse shaping techniques. In the optics regime, pulse shaping techniques such as spatial ligth modulation \cite{PeEr2005, Lukens2013}, eletro-optic modulation \cite{Kolchin2008, Specht2009, Olislager2010, Zhang2012, Agha2014}, cross-phase modulation \cite{Matsuda2016,Keller2004} have been demonstrated. In the microwave regime, arbitrary waveform generator and pulse shaping by tunable resonator-emitter coupling can be applied \cite{Srinivasan2011, Pechal2014, Forn-Diaz2017}. The shaped photon pulse is then input into the waveguide from both directions and it can drive the atomic transition between states $|b\rangle$ and $|a\rangle$. At a predetermined time, we apply a  classical $\pi$ pulse to transfer the population in state $|a\rangle$ to state $|c\rangle$. Since state $|c\rangle$ has a very slow decay rate, the prepared quantum state can then be preserved for an extended period of time.

The interaction Hamiltonian of the system in the rotating wave approximation is given by  \cite{Liao2015, Liao2016a}
\begin{equation}
H=\hbar\sum_{j=1}^{N}\Big [\sum_{k}g_{k}e^{ikr_{j}}a_{k}\sigma_{j}^{+}e^{-i\delta \omega_{k}t}+\Omega(t)|a\rangle_{j}\langle c|+H.c.\Big ]
\end{equation}
The first term is the coupling between the waveguide photon and the $|a\rangle \leftrightarrow |b\rangle$ transition with coupling strength $g_{k}=\boldsymbol{\mu}_{ab}\cdot\mathbf{E}_{k}(\mathbf{r}_j)/\hbar$ ( $\boldsymbol{\mu}_{ab}$ is the transition dipole moment, $\mathbf{E}_{k}(\mathbf{r}_j)$ is the guided photon field strength at position $r_j$, and $\hbar$ is the Planck constant). Since the phase of $g_{k}$ only affects the overall phase of the quantum state, we can safely assume that $g_{k}$ is a real number (i.e., $g_{k}=g_{k}^{*}$). The second term is the coupling between the classical driving light and the $|a\rangle \leftrightarrow |c\rangle$ transition with Rabi frequency $\Omega(t)$. Here, $a_{k}^{\dagger } (a_{k}^{-})$  are the creation (annihilation) operators of the waveguide photon modes with wavevector $k$, and $\sigma_{j}^{+}=|a\rangle_j\langle b| ~ (\sigma_{j}^{-}=|b\rangle_j\langle a|)$ is the raising (lowering) operator of the $j$th emitter for the $|a\rangle \leftrightarrow |b\rangle$  transition. $\delta\omega_{k}=(|k|-k_{a})v_{g}$ is the detuning between the photon frequency and the atomic transition frequency where $k_{a}$ is the wave vector at frequency $\omega_{a}$ and $v_{g}$ is the group velocity \cite{Shen2005a}.

For a single photon excitation, the quantum state of the system at time $t$ can be expressed as 
\begin{equation}
|\Psi(t)\rangle =\sum_{j=1}^{N}[a_{j}(t)|a_{j},0\rangle+c_{j}(t)|c_{j},0\rangle]+\sum_{k}\beta_{k}(t)|b,1_{k}\rangle 
\end{equation}
where $|a_{j},0\rangle ~(|c_{j},0\rangle)$ is the state that the $j$th atom is in the excited state $|a\rangle ~(|c\rangle)$ while the other atoms are in the ground state $|b\rangle$ with zero photon in the waveguide, $|b,1_{k}\rangle$ is the state that all the atoms are in the ground state and one photon with wavevector $k$ is in the waveguide mode. From the Schr\"{o}dinger equation $i\hbar \partial /\partial_{t} |\psi(t)\rangle = H |\psi(t)\rangle$, we have
\begin{align}
\dot{a_{j}}(t)=&-i\sum_{k}g_{k}e^{ikr_{j}-i\delta\omega_{k}t}\beta_{k}(t) -i\Omega(t)c_{j}(t),  \\
\dot{\beta}_{k}(t)=&-i\sum_{j=1}^{N_a}g_{k}^{*}e^{-ikr_{j}}e^{i\delta\omega_{k}t}a_{j}(t), \\
\dot{c_{j}}(t)= & -i\Omega^{*}(t)a_{j}(t).
\end{align}
Formally integrating Eq. (4) and inserting the result into Eq. (3) and using the Weisskopf-Wigner approximation, we can obtain the dynamics of the emitters which is given by \cite{Liao2015}
\begin{equation}
\dot{a}_{j}(t)= b_{j}(t)-\sum_{l=1}^{N}(\frac{\Gamma}{2}e^{ik_{a}r_{jl}}-\gamma\delta_{jl})a_{l}(t-\frac{r_{jl}}{v_{g}})-i\Omega(t)c_{j}(t),  
\end{equation}
where $b_{j}(t)=-\frac{i}{2\pi}\sqrt{\frac{\Gamma v_{g}L}{2}}\int_{-\infty }^{\infty }\beta_{k}(0)e^{ikr_{j}-i\delta \omega_{k}t}dk$  is the excitation by the incident photon with $\beta_{k}(0)$ being the spectrum of the incident photon. $\Gamma=2L|g_{k_a}|^2/v_{g}$ is the decay rate due to the waveguide photon modes, $\gamma$ is the decay rate to the free space, and $r_{jl}$ is the distance between the $j$th and $l$th emitters. 

From Eqs. (5) and (6) we can calculate the emitter excitation as a function of time. The photon spectrum in arbitrary time can be then calculated by integrating Eq. (4). For the purpose of quantum state preparation here, we consider the emission spectrum when the emitter system is prepared in a specific quantum state (i.e., $|\psi\rangle=\sum_{j=1}^{N}a_{j}(0)|g\cdots e_j \cdots g\rangle$) with no photon input (i.e., $\beta_{k}(0)=0$). In this case, the photon spectrum at $t\rightarrow \infty$ is given by
\begin{equation}
\beta_{k}(t)=-i\sqrt{\frac{\Gamma v_{g}}{2L}}\sum_{j=1}^{N_a}e^{-ikr_j}\chi _{j}(k),
\end{equation}
where $\chi _{j}(k)=\int_{-\infty}^{\infty}e^{i\delta\omega_k t'}a_j(t')\Theta(t')dt'$
with $\Theta(t')=1$ for $t'\geqslant 0$ and $\Theta(t')=0$ for $t'<0$. By performing the inverse Fourier transformation 
$a_{j}(t)\Theta(t)=\frac{1}{2\pi}\int_{-\infty}^{\infty}\chi _{j}(k)e^{-i\delta\omega_{k} t}d\delta \omega_{k}$,
and using the relation 
$\frac{d}{dt}[a_{j}(t)\Theta(t)]=\dot{\alpha}_{j}(t)\Theta(t)+a_{j}(t)\delta (t)$,
we have from Eq. (6)
\begin{equation}
(-i\delta\omega_{k})\chi_{j}(k)=-\frac{\Gamma}{2}\sum_{l=1}^{N_a}e^{i(k_a+\delta k)r_{jl}}\chi_{l}(k)+a_{j}(0)
\end{equation}
whose solution is 
\begin{equation}
\chi_{j}(\delta k)=\sum_{l=1}^{N}[M(\delta k)]_{jl}^{-1}a_{l}(0)
\end{equation}   
where $M(\delta k)$ is an $N\times N$ matrix with matrix element given by $[M(\delta k)]_{jl}=\frac{\Gamma}{2}e^{i(k_a+\delta k)r_{jl}}-i\delta\omega_{k}\delta_{jl}$. Hence, the emission spectrum is given by \cite{Liao2016a}  
\begin{equation}
\beta_{k}=-i\sqrt{\frac{\Gamma v_{g}}{2L}}\sum_{j,l=1}^{N}a_{l}(0)[M^{-1}(\delta k)]_{jl}e^{-ikr_j},
\end{equation}
where $M^{-1}(\delta k)$ is the matrix inverse of $M(\delta k)$.

According to our theory, to prepare a single-excitation quantum state $|\psi\rangle=\sum_{j=1}^{N}a_{j}(0)|g\cdots e_j \cdots g\rangle$ in an $N$ emitter system, the spectrum of the incident photon pulse should be the complex conjugate of the emission spectrum of the system with Hamiltonian $H^{*}(-t)$ which is initially prepared in state $|\psi^{*}\rangle$.  Since the only difference between $H(t)$ and $H^{*}(-t)$ is that $k\rightarrow -k$, the spectrum emitted by a quantum state $|\psi^{*}\rangle=\sum_{j=1}^{N}a_{j}^{*}(0)|g\cdots e_j \cdots g\rangle$ under $H^{*}(-t)$ is given by Eq. (10) with $a_{l}(0)\rightarrow a_{l}^{*}(0)$ and $e^{-ikr_j}\rightarrow e^{ikr_j}$ and its complex conjugate is given by 
\begin{equation}
\beta_{k}^{*}=i\sqrt{\frac{\Gamma v_{g}}{2L}}\sum_{j,l=1}^{N}a_{l}(0) [M^{-1}(\delta k)]_{jl}^{*}e^{-ikr_j}.
\end{equation}
By injecting a single photon with spectrum shown in Eq. (11), the emitters can be driven to the quantum state $|\psi\rangle=\sum_{j=1}^{N}a_{j}(0)|g\cdots e_j \cdots g\rangle$.

In the following, we show how to prepare the desired quantum state onto the emitter system by designing specific $\beta_{k}(0)$. For simplicity, we maily consider the case when the loss to the free space modes is negligible ($\gamma=0$), which is reasonable because near-perfect waveguide with very small $\gamma$ has  been reported \cite{Lund-Hansen2008, Quan2009}. Nonetheless, our numerical simulation shows that even if $\gamma$ is non-negligible our scheme still works well when $\gamma$ is not very large. We first neglect the classical light pulse and show how the desired quantum state in the $|a\rangle$ and $|b\rangle$ subspace can be prepared. Then we show how to drive the system into a more stable state by applying a classical $\pi$ pulse on the $|a\rangle\leftrightarrow |c\rangle$ transition.

\begin{figure*}
\includegraphics[width=1.9\columnwidth]{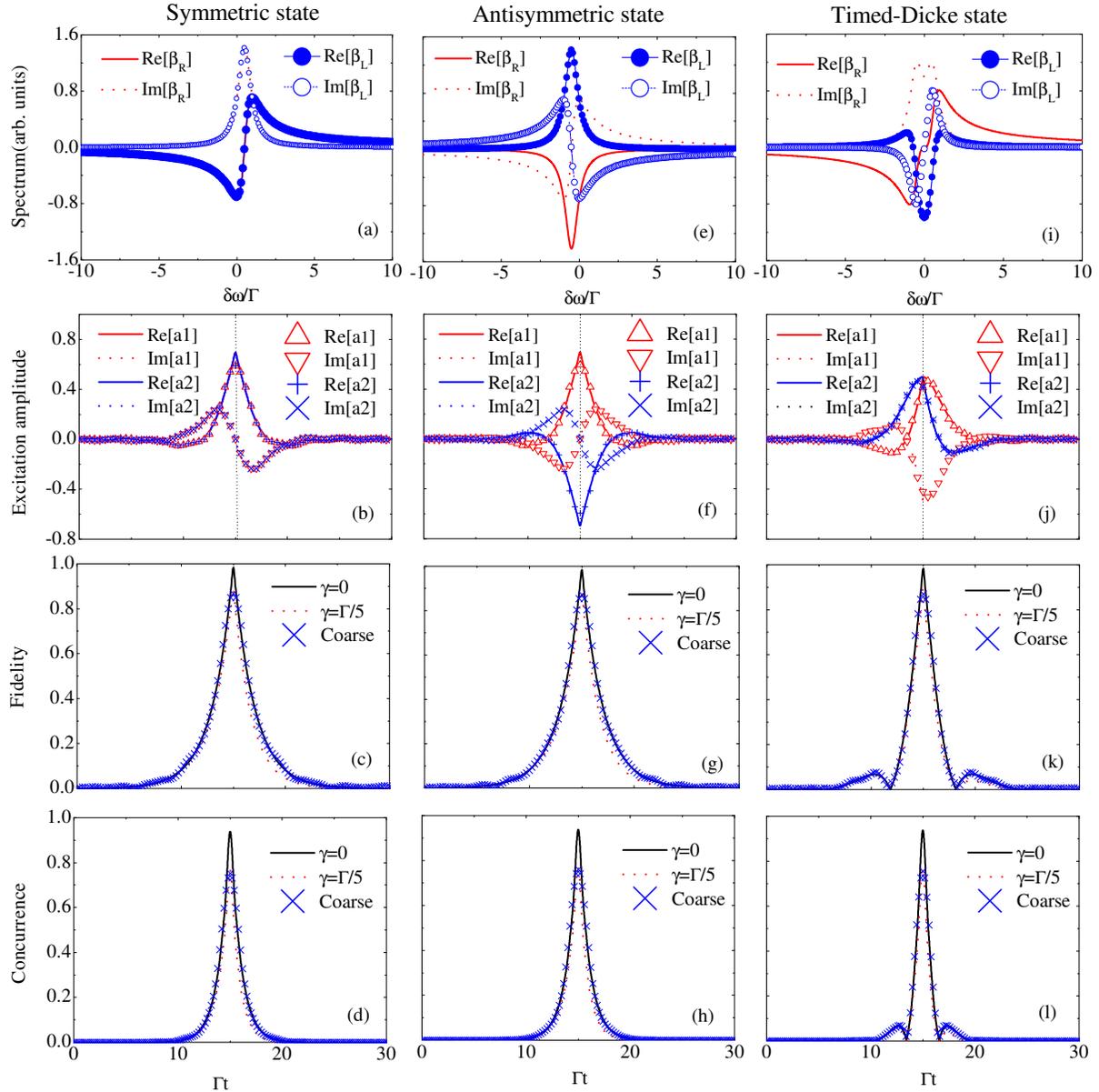}
\caption{(Color online) Two-emitter excitation for three different single-photon pulses input. (a-d) Symmetric state preparation. (e-h) Antisymmetric state preparation. (i-l) Timed-Dicke state preparation. The first row is the spectrum of the incident photon (solid lines: real part; dashed lines: imaginary part). The second row is the emitter excitation amplitude as a function of time (solid lines: real part; dashed lines: imaginary part; symbols: coarse-grained sampling).  The third (fourth) row is the fidelity (concurrence) as a function of time where the solid line is the result with $\gamma=0$, the dashed line is the result with $\gamma=\Gamma/5$, and the blue cross symbol is the result with coarse-grained sampling. The distance between the two emitters is $0.25\lambda$ and $t_0=15 v_g/\Gamma$. } 
\end{figure*}

\section{Numerical examples}

In this section, we numerically demonstrate how to prepare a desired quantum state onto the emitter system by carefully designing the incident photon pulse.

\subsection{Two-emitter case}

Let us first look at the simplest two-emitter system. Supposing that  $|\psi\rangle=a_1 |eg\rangle + a_{2}|ge\rangle$ to be prepared, the required photon spectrum can be calculated from Eq. (11) with
\begin{equation}
M(\delta k)=\begin{bmatrix}
\frac{\Gamma}{2}-i\delta\omega_{k}, & \frac{\Gamma}{2}e^{i(k_a+\delta k)d} \\
\frac{\Gamma}{2}e^{i(k_a+\delta k)d}, &  \frac{\Gamma}{2}-i\delta\omega_{k}
\end{bmatrix}
\end{equation}
where $d=r_{12}$ is the distance between the two emitters. 
For example, to prepare a symmetric state $|\psi_{S}\rangle=\frac{1}{2}(|eg\rangle+|ge\rangle)$, we can inject a photon with spectrum given by Eq. (11) with $a_1=a_2=1/\sqrt{2}$ and it is shown in Fig. 2(a). The photon pulse is assumed to be $v_{g}t_{0}$ away from the first emitter with $t_{0}=15/\Gamma$ in the numerical calculations in this subsection. The left and right propagating photon spectrum have the same magnitude and phase. The emitter excitation amplitudes as a function of time for this input are shown in Fig. 2(b). The two emitters are excited and deexcited at the same pace and at time $t=15/\Gamma$, $\text{Re}[a_1]=\text{Re}[a_{2}]=0.7\approx 1/\sqrt{2}$ and $\text{Im}[a_{1}]=\text{Im}[a_{2}]=0$ which indicates that the symmetric state $|+\rangle=(|eg\rangle + |ge\rangle)/\sqrt{2}$ has been successfully prepared. In practice, it is very difficult to shape the photons perfectly. We also numerically calculate the case when the required spectrum is coarse-grained sampling.  Assuming that the input spectrum consists of 20 discrete frequency components uniformly distributed from $-2.5\Gamma$ to $2.5\Gamma$, the emitter excitation amplitudes are shown as the symbols in Fig. 2(b) from which we see that the excitation follows the curve very well. Figure 2(c) shows the fidelity between the evolving emitter state and the symmetric state as a function of time. When $\gamma=0$ (black solid curve), the fidelity can approach unit  at $t=15/\Gamma$ which indicates that the symmetric state has been successfully prepared. Even if $\gamma$ is non-negligible (e.g., $\gamma=\Gamma/5$, red dashed line) or the input spectrum is coarse-grained sampled (blue cross symbol), the fidelity can be still about $90\%$. These results show that even if the waveguide or the input spectrum are imperfect, our scheme can still work well. To show that the entanglement has been created in this process, in Fig. 2(d) we show the concurrence as a function of time for three different cases as in Fig. 2(c). The concurrence as a function of time in this two-emitter example is given by $2|a_{1}(t)||a_{2}(t)|$ \cite{Liao2016}. It is clearly seen that entanglement has been created between this preparation process. When $\gamma=0$, the concurrence at $t=15/\Gamma$ is about $0.94$. The concurrence when $\gamma=\Gamma/5$ is about $0.76$ and it is about $0.75$ for the coarse sampling case.

Similarly, if the incident photon has a spectrum given by Eq. (11) but with $a_{1}=-a_{2}=1/\sqrt{2}$, we can prepare the emitter system into an antisymmetric state $|-\rangle=(|eg\rangle - |ge\rangle)/\sqrt{2}$. The corresponding incident photon spectrum is shown in Fig. 2(d). Different from the symmetric case, the left and right propagating spectra in the antisymmetric case have opposite phase. Figure 2(e) shows the emitter excitation amplitude as a function of time. The two emitters have the same excitation magnitude but opposite phase.  At $t=15/\Gamma$,  $\text{Re}[a_{1}]=-\text{Re}[a_{2}]=0.7\approx 1/\sqrt{2}$, and $\text{Im}[a_{1}]=\text{Im}[a_{2}]\simeq 0$. The fidelity with respect to the antisymmetric state is about $98\%$ which clearly shows that the antisymmetric state has been successfully prepared (back solid line in Fig. 2(f)). The fidelity with $\gamma=\Gamma/5$ and the coarse-grained spectrum can be still be about $90\%$. Similar to the symmetric case, entanglement is also clearly present which is shown in Fig. 2(h).

\begin{figure*}
\includegraphics[width=0.48\columnwidth]{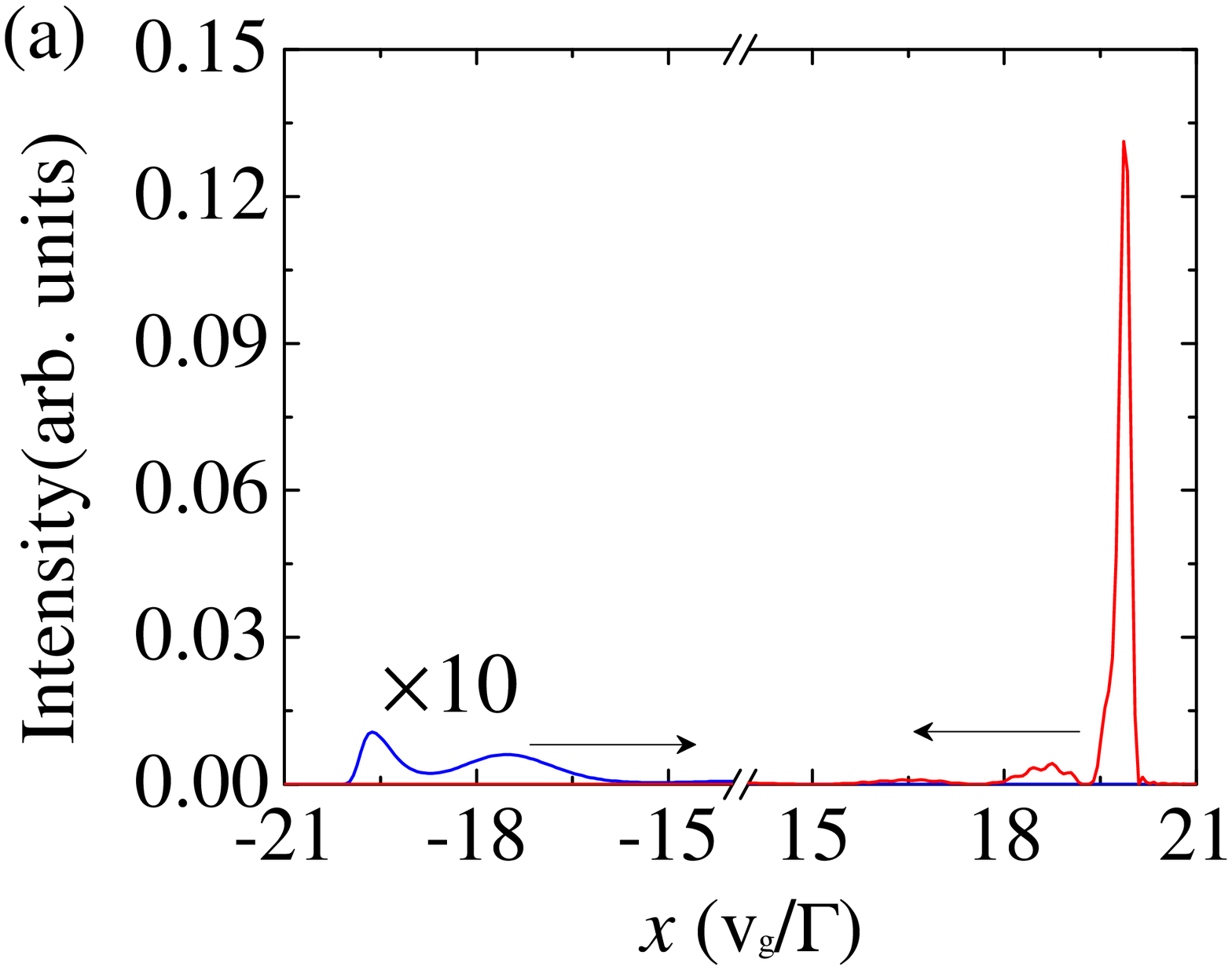}
\includegraphics[width=0.48\columnwidth]{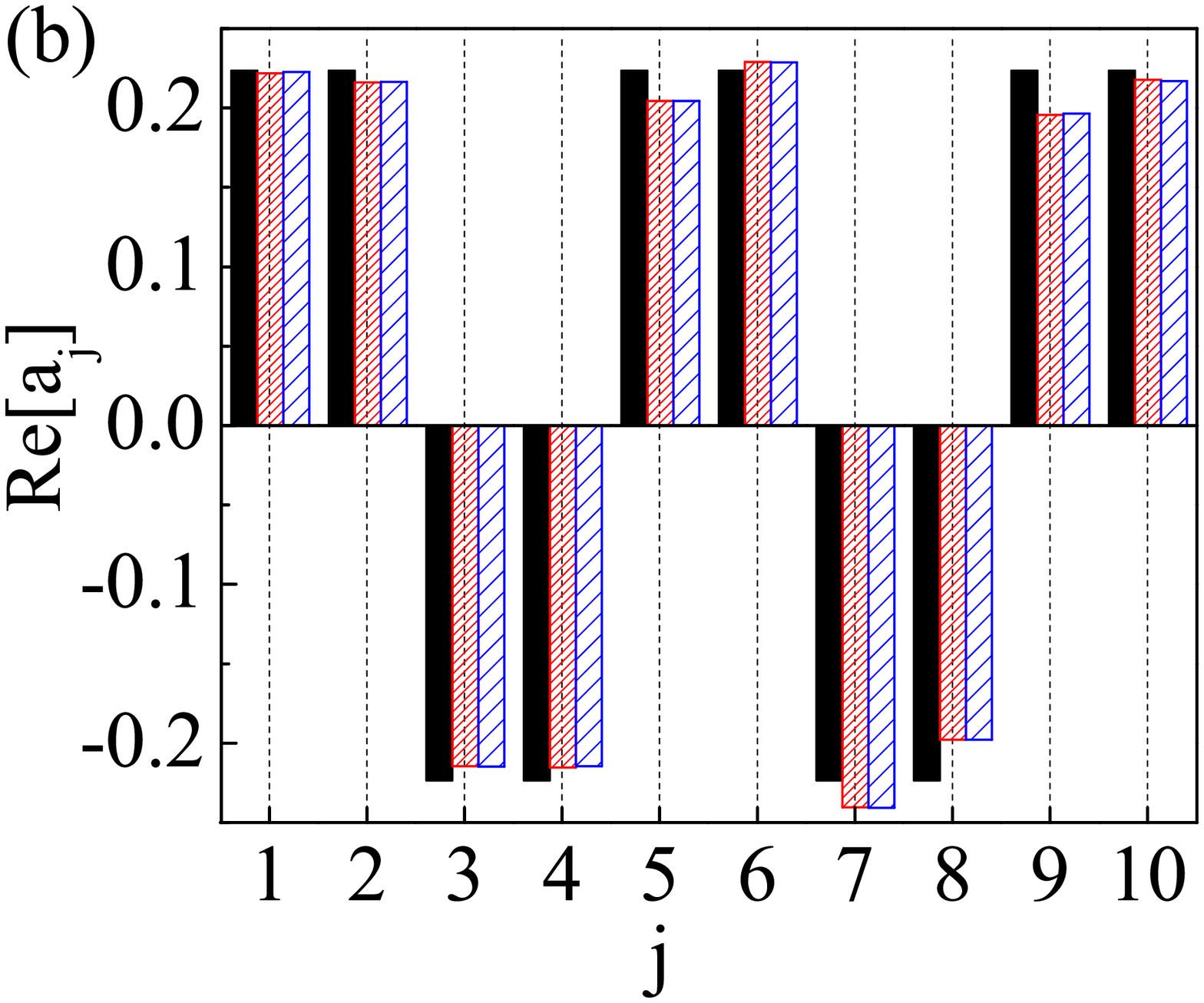}
\includegraphics[width=0.48\columnwidth]{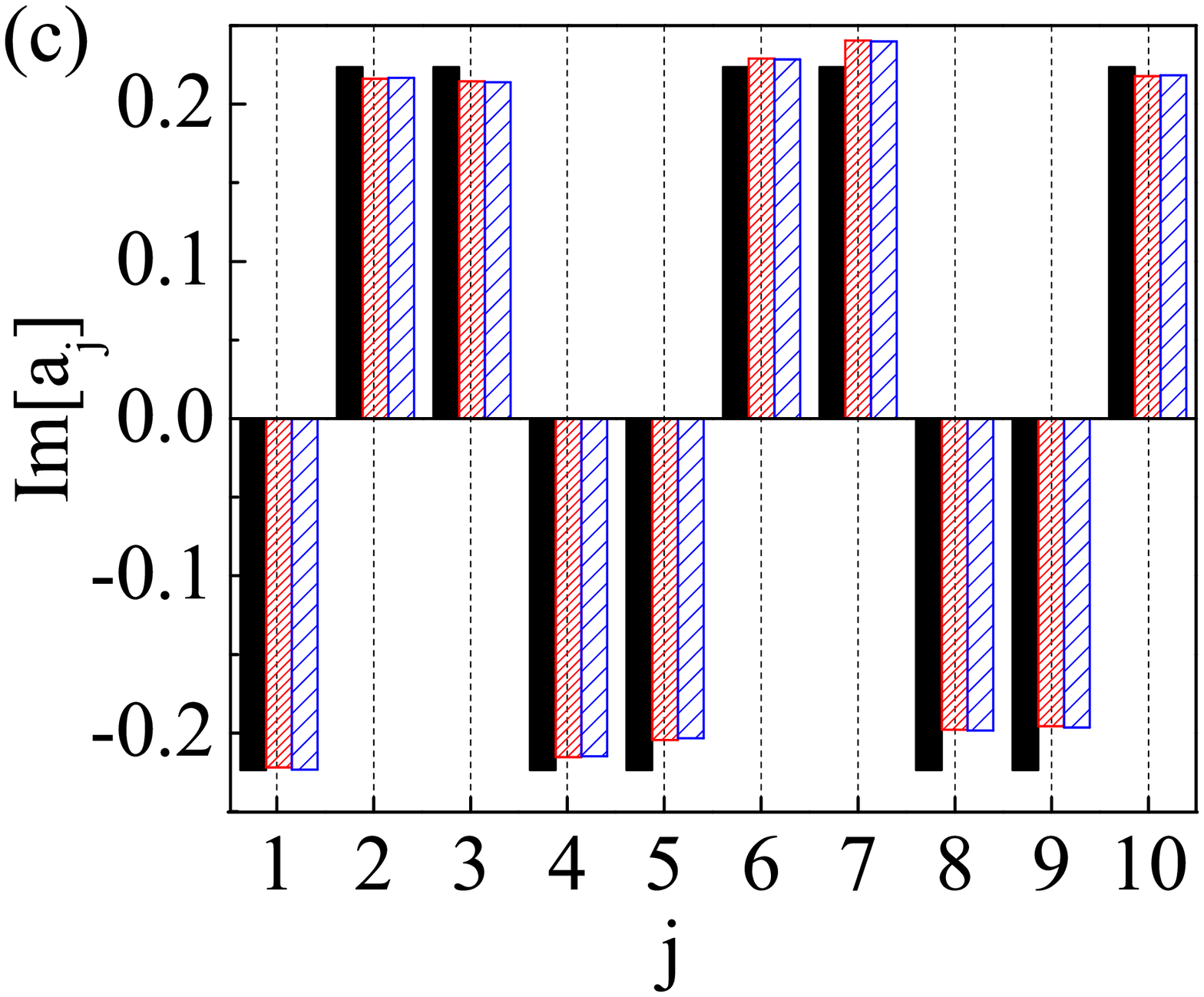}
\includegraphics[width=0.48\columnwidth]{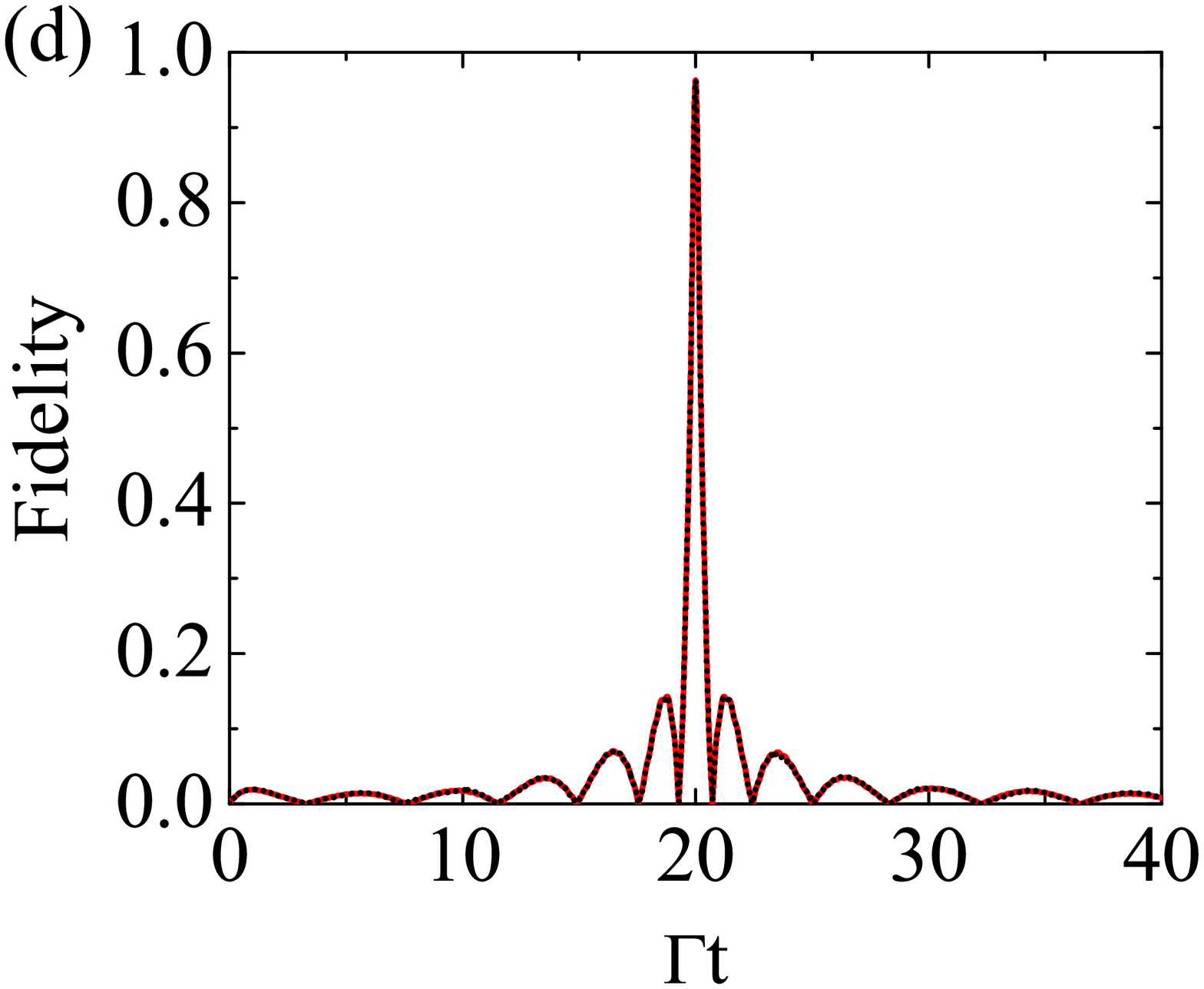}
\caption{(Color online) Preparation of timed-Dicke state in a ten-emitter system ($t_{0}=20/\Gamma$). (a) The intensity (arbitrary units) of the incident pulse as a function of position.  The real (b) and imaginary (c) parts of the excitation amplitude for each emitter at $t=20/\Gamma$.  The bars with solid filling are the expected excitation amplitude, the red bars with dense pattern are the numerical calculated excitation amplitudes by the incident pulse shown in (a), and the bars with sparse pattern are  the numerical results when the incident spectrum and the emitter positions have $10\%$ uncertainties.  (d) The fidelity as a function of time. The red solid line is the fidelity without uncertainty and the black dotted line is the fidelity with $10\%$ uncertainty on the incident spectrum and the emitter positions.} 
\end{figure*}

In addition to the symmetric and antisymmetric states, we can also drive the system to the interesting timed-Dicke state \cite{Scully2006}.  The timed-Dicke state can be probabilistically prepared by delayed-choice measurent \cite{Scully2006} and it may be used for ultrasensitive quantum metrology \cite{Wang2014a}. However, deterministic preparation of the timed-Dicke state is still an open question. In our method, to prepare a timed-Dicke state, we can input a single photon with a spectrum given by Eq. (11) with $a_{1}=e^{ik_{a}r_{1}}$ and $a_{2}=e^{ik_{a}r_{2}}$. For example, if the two emitters have separation $\lambda/4$ with $r_{1}=-\lambda/8$ and $r_{2}=\lambda/8$, the corresponding timed-Dicke state $|\psi_{TD}\rangle=(e^{-i\pi/4}|eg\rangle +e^{i\pi/4}|ge\rangle)/\sqrt{2}$ with $a_1=e^{-i\pi/4}$ and $a_2=e^{i\pi/4}$. The desired photon spectrum is shown in Fig. 2(g).  The left and right propagating modes have different spectrum amplitudes and phases. By injecting the single photon with spectrum shown in Fig. 2(g), the emitter excitation amplitude as a function of time is shown in Fig. 2(h). At $t=15/\Gamma$, $\text{Re}[a_{1}]=\text{Re}[a_{2}]=\text{Im}[a_{2}]=0.492$ and $\text{Im}[a_{1}]=-0.492$. This state has a fidelity with respect to the timed-Dicke state about $98.5\%$ which infers that the required timed-Dicke state is successfully prepared by a single photon pulse. For imperfect waveguide such that $\gamma=\Gamma/5$, the fidelity to prepare the timed-Dicke state with the same pulse can still be about $90\%$. With coarse-grained sampling, the fidelity can still be about $87\%$. The concurrence dynamics is similar to the fidelity curve and the entanglement can approach maximum which is shown in Fig. 2(l).

\subsection{Multiple-emitters case}

This method can also be applied in a quantum system with multiple number of emitters. The maximum number of emitters we can manipulate is limited by the bandwidth of the waveguide. The collective emission linewidth of the emitters should be less than the bandwidth of the waveguide.   For example, to prepare a timed-Dicke state on a ten-emitter system, i.e., $|\psi_{TD}\rangle=1/\sqrt{10}\sum_{j=1}^{10}e^{ik_a r_j}|g\cdots e_j \cdots g\rangle$ with $r_j=(-1.125+0.25i)\lambda$,  the required spectrum is given by Eq. (11) with $a_j=e^{i(2j-1)\pi/4}/\sqrt{10}$ and the corresponding pulse shape in the real space is shown in Fig. 3(a). The red curve is the photon pulse incident from the right and the blue curve (is amplified by ten times for clarity) is that from the left. It is clearly seen that most of the photon pulse is coming from one direction which is the demonstraction of the directional emission of the timed-Dicke state \cite{Scully2006}. On applying this pulse, the real part and imaginary part of the excitation amplitude for each emitter  at $t=20/\Gamma$  are shown in Fig. 3(b) and 3(c), respectively. The bars with black solid filling are the expected excitation amplitudes and the bars with filling patterns are the numerical results which shows that the generated quantum state (red bars with dense pattern) by the incident pulse shown in Fig. 3(a) is very close to the expected time-Dicke state. Even if the incident spectrum and the emitter positions have $10\%$ uncertainty, the numerical excitation amplitudes (blue bars with sparse pattern) are still very close to the expected quantum state. This can be also seen from Fig. 3(d) where the fidelity between the generated state and the expected state is shown. The red solid curve is the fidelity when the incident pulse is shown in Fig. 3(a) and the emitter positions are precisely determined, while the black dotted curve is the fidelity when noise is included, i.e., the incident pulse and  the emitter positions have $10\%$ uncertainty. At $t=20/\Gamma$, both of them can have maximum fidelity about $96\%$ which shows that the timed-Dicke state can be deterministically prepared and it is very robust against the noises like the imperfect pulse shaping and the emitter position uncertainty.

\begin{figure}
\includegraphics[width=0.9\columnwidth]{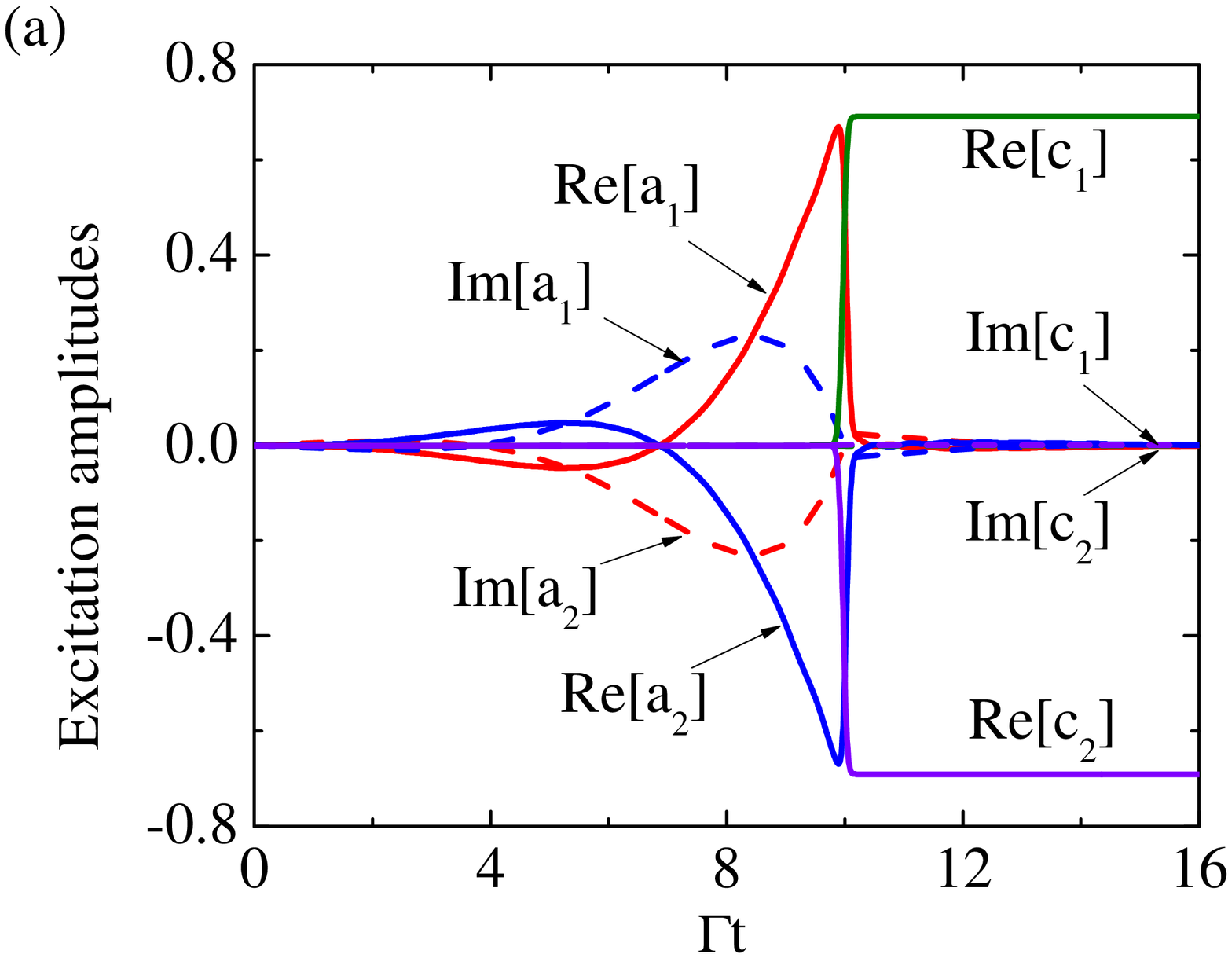}
\includegraphics[width=0.9\columnwidth]{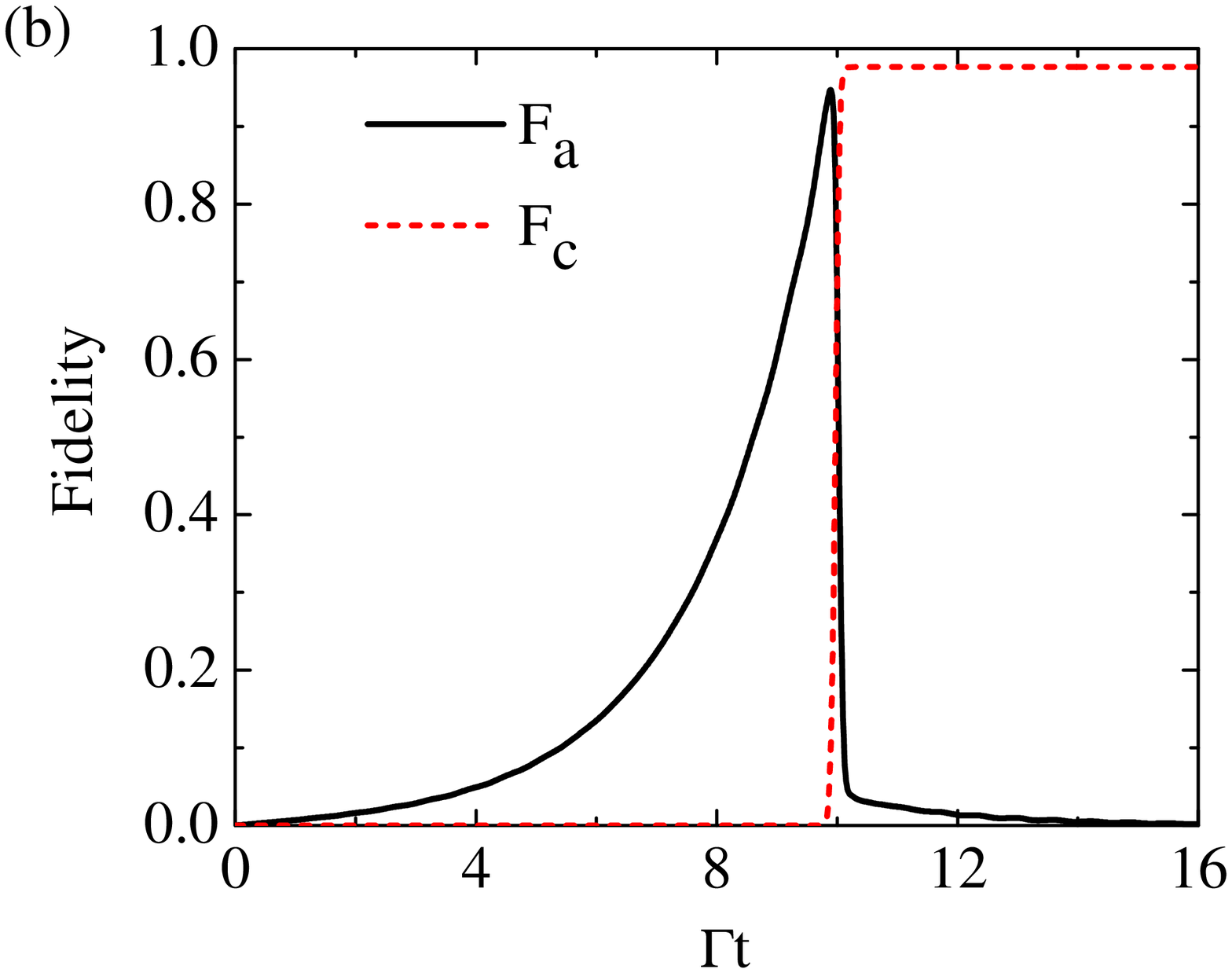}
\caption{(Color online) Robust quantum state preparation. (a) The two-emitter excitation as a function of time. (b) The fidelity with respect to the symmetric state as a function of time. $t_{0}=10/\Gamma$. } 
\end{figure}

\section{Robust quantum state preparation}

In the previous discussions, we have shown that a desired quantum state can be successfully prepared at certain time. However, after that the quantum state decay quickly. Here, we show how to preserve these quantum states in a more robust quantum state via Raman process. Raman process has been used to create quantum entanglement between two cascaded atom-cavity system \cite{DiFidio2008}. Our idea here is that we can apply a $\pi$ pulse to transfer the population in the state $|a\rangle$ to a more stable state $|c\rangle$ right after the first pulse. For example, to prepare a antisymmetric state in a two-emitter system, we can apply a single photon pulse to excite the transition from $|b\rangle$ to $|a\rangle$ and the photon spectrum is given by Eq. (11) with $a_1=-a_2=1/\sqrt{2}$. Meanwhile, we apply a classical light pulse with Rabi frequency $\Omega(t)=\frac{\sqrt{\pi}}{2\delta}e^{-(t-t_0)^2/\delta^2}$ and $\delta=0.1\Gamma$ to induce a transition from state $|a\rangle$ to state $|c\rangle$. Since $\int_{-\infty}^{\infty}\Omega(t)dt=\pi/2$, the pulse can completely transfer the population at state $|a\rangle$ to state $|c\rangle$. The emitter excitation amplitudes as a function of time are shown in Fig. 4(a) from which we can see that the excitation in state $|a\rangle$ can be transferred to the state $|c\rangle$ when the classical pulse is applied at around $t=10/\Gamma$. The fidelity $F_{a}$ between the state $a_1(t)|a_{1}b_{2}\rangle+a_{2}(t)|b_{1}a_{2}\rangle$ and the antisymmetric state $(|a_{1}b_{2}\rangle-|b_{1}a_{2}\rangle)/\sqrt{2}$ is shown as the black solid curve in Fig. 4(b), and the fidelity $F_{c}$ between the state $c_1(t)|c_{1}b_{2}\rangle+c_{2}(t)|b_{1}c_{2}\rangle$ and the antisymmetric state $(|c_{1}b_{2}\rangle-|b_{1}c_{2}\rangle)/\sqrt{2}$ is shown as the red dashed curve in Fig. 4(b). As the single photon pulse is applying, $F_a$ is increasing and approaching about 1 at about $t=10/\Gamma$. Then we apply a classical $\pi$ pulse, $F_a$ decreases very quickly while $F_{c}$ increases rapidly and approaches about 1. The antisymmetric state is therefore successfully prepared in a more robust state because the state $|c\rangle$ is a substable state.

\section{Summary}

In summary, we have shown that driven by a shaped photon pulse a desired quantum state such as the Dicke and timed-Dicke states in a 1D waveguide-QED system can be sucessfully prepared with fidelity approaching unit. This method is based on the time-reverse of a quantum emission process. The design of the preparation process is straightforward and the number of photon required is minimized. We also propose a method to transfer the prepared quantum state to a more robust state by applying a classical $\pi$ pulse. The method shown here can be experimentally demonstrated in the circuit-QED system where strong coupling and single microwave photon pulse shaping have been successfully achieved \cite{Pechal2014,Forn-Diaz2017, Xiang2013}. This work may find important applications in the waveguide-QED-based quantum science.  In this work, we mainly consider the single-excitation case which is more tractable. However, this method should be able to be extended to the more general case where multiple excitations can occur. 

\section*{Acknowledgment}
We thank S. -W. Li, J. You and X. Zeng for helpful discussions. This work is supported by a grant from the Qatar National Research Fund (QNRF) under the NPRP project 8-352-1-074.

%\bibliography{/Users/zeyangliao/research/ref/ref}

\end{document}